\documentclass{article}
\usepackage{spconf,amsmath,graphicx}
\usepackage{epsfig,graphics,epsf,color,amsmath,balance,cite,amsfonts,multirow,stfloats,algorithm,algorithmic,mathrsfs,amssymb,color,subfigure}
\usepackage[dvipdfm,unicode,colorlinks, linkcolor=black,anchorcolor=black,citecolor=black]{hyperref}
\newtheorem{theorem}{Theorem}
\newtheorem{lemma}{Lemma}
\newtheorem{proposition}{Proposition}

\newcommand*{\QEDB}{\hfill\ensuremath{\square}}

\title{INTEGRATED SENSING AND FULL-DUPLEX COMMUNICATION: JOINT TRANSCEIVER BEAMFORMING AND POWER ALLOCATION}

\name{Zhenyao He$^{\star}$, Wei Xu$^{\star}$, Hong Shen$^{\star}$, Derrick Wing Kwan Ng$^{\dagger}$, Yonina C. Eldar$^{\ \ddagger}$, and Xiaohu You$^{\star}$
}

\address{$^{\star}$ National Mobile Communications Research Laboratory, Southeast University\\
$^{\dagger}$ School of Electrical Engineering and Telecommunications, University of New South Wales\\
$^{\ddagger}$ Faculty of Mathematics and Computer Science, Weizmann Institute of Science}

\begin{document}

\maketitle
\begin{abstract}
Beamforming design has been widely investigated for integrated sensing and communication (ISAC) systems with full-duplex (FD) sensing and half-duplex (HD) communication. To achieve higher spectral efficiency, in this paper, we extend existing ISAC beamforming design by considering the FD capability for both radar and communication. Specifically, we consider an ISAC system, where the base station (BS) performs target detection and communicates with multiple downlink users and uplink users reusing the same time and frequency resources. We jointly optimize the downlink dual-functional transmit signal and the uplink receive beamformers at the BS and the transmit power at the uplink users.
The problem is formulated to minimize the total transmit power of the system while guaranteeing the communication and sensing requirements.
The downlink and uplink transmissions are tightly coupled, making the joint optimization challenging.
To handle this issue, we first determine the receive beamformers in closed forms with respect to the BS transmit beamforming and the user transmit power and then suggest an iterative solution to the remaining problem.
We demonstrate via numerical results that the optimized FD communication-based ISAC leads to power efficiency improvement compared to conventional ISAC with HD communication.
\end{abstract}
\begin{keywords}
Integrated sensing and communication (ISAC), full-duplex (FD) communication, joint optimization.
\end{keywords}
\section{Introduction}
\label{sec:intro}

In future networks, a pressing problem is to satisfy the heterogeneous requirements of reliable sensing and efficient communication among the wireless terminals \cite{W.XuJSTSP2022}.
Integrated sensing and communication (ISAC) has become an appealing technique to meet this requirement and has attracted considerable research interest. It has been shown in the literature \cite{overviewJ.Zhang,J.Zhang2022CSTEnabling,overviewF.Liu} that ISAC can significantly enhance the spectral efficiency and reduce implemental cost by sharing spectral resources and reusing expensive hardware architectures. Also, joint design of communication and sensing can help improve the performances of both functionalities.

Many works have studied transmit design in multi-antenna ISAC systems by focusing on transmit beamforming optimization, e.g., \cite{F.LiuTWC2018,XLiuTSP2020,Z.HeWCL2022}.
By considering the radar echo reception concurrently, joint transceiver design in ISAC systems were further investigated,
e.g., in \cite{Liu2022CRB,L.Chen2022JSAC,M.Temiz2021TCCN,X.Wang2022CL}.
In these works, the radar receiver operates simultaneously while transmitting, i.e., in a full-duplex (FD) manner.
With FD radar, however, the integrated communication functionality occurs only in either the downlink transmission \cite{Liu2022CRB,L.Chen2022JSAC} or the uplink transmission \cite{M.Temiz2021TCCN,X.Wang2022CL}, operating in a half-duplex (HD) manner.
Therefore, it is natural to consider the FD capability also for communication to achieve higher spectral efficiency.
Under this setup, there is not only interference between sensing and communication functionalities, but also coupling between uplink and downlink transmissions, that significantly complicate the ISAC design.
Existing algorithms in \cite{Liu2022CRB,L.Chen2022JSAC,M.Temiz2021TCCN,X.Wang2022CL} cannot be straightforwardly applied to address these challenges.
Specifically, the algorithms designed in \cite{Liu2022CRB,L.Chen2022JSAC} do not incorporate the impact of uplink communication. In \cite{M.Temiz2021TCCN,X.Wang2022CL}, only a pure downlink sensing signal is sent and the uplink transmit power is fixed, without considering the possibility of downlink communication nor designing the uplink transmission.

In this paper, we extend existing ISAC beamforming design to a general case by considering the FD capability for both radar and communication. Specifically, we consider that a BS receives and transmits signals from multiple uplink users and downlink users reusing the same time and frequency resources. The downlink transmit signal is an ISAC signal applied for both conveying information to the downlink users and performing a sensing task of point target detection. The BS also simultaneously conducts uplink communication signal reception and processes the radar echo signal. Our goal is to jointly design the transceiver beamforming at the BS and the transmit power at the single-antenna uplink users.

To respectively detect both the sensing target and multiuser uplink signals with low complexity from the received signal at the BS, multiple linear receive beamformers are employed. Then, the corresponding radar and uplink communication signal-to-interference-plus-noise ratios (SINRs) are mathematically obtained. We formulate a power minimization problem for the joint optimization, by constraining the minimal SINR requirements of target detection, uplink communications, and downlink communications.
We first determine the optimal receive beamformers to maximize the SINR of target detection and the SINRs of uplink communications, respectively, which are derived in closed forms with respect to the downlink transmit signal and the uplink transmit power.
Then, we invoke the closed-form receivers to obtain an equivalent problem and propose an iterative algorithm to solve it by applying the techniques of rank relaxation and successive convex approximation (SCA).
Simulation results show that our optimized FD communication-based ISAC has much higher power efficiency than the previous frameworks that integrated sensing with HD communication.

The rest of this paper is organized as follows. In Section~\ref{sec:2}, we present the system model and the problem formulation.
In Section~\ref{sec:3}, we propose an efficient algorithm for the considered problem. Numerical simulations are provided in Section~\ref{sec:4} and conclusions are drawn in Section~\ref{sec:5}.

\section{SYSTEM MODEL AND PROBLEM FORMULATION}
\label{sec:2}
A dual-functional FD BS equipped with $N_t$ transmit antennas and $N_r$ receive antennas receives the communication signals from $K$ uplink users and sends a downlink ISAC signal via the same time-frequency resource.
The downlink ISAC signal is used for simultaneously communicating with $L$ downlink users and performing target detection on a point radar target.

\subsection{Signal Model}
Following \cite{XLiuTSP2020,Z.HeWCL2022,Liu2022CRB}, the downlink transmit ISAC signal, denoted as $\mathbf x \in \mathbb C^{N_t \times 1}$, can be expressed as
\begin{align}\label{signal:xISAC}
\mathbf x = \sum_{l=1}^L \mathbf v_l s_l + \mathbf s_0,
\end{align}
where $\mathbf v_l \in \mathbb C^{N_t \times 1}$ stands for the beamforming vector associated with downlink user $l$, $l \in \{1,\cdots,L\},$ and $s_l \in \mathbb C$ is the data symbol of user $l$ with unit power, i.e., $\mathbb{E}\{ |s_l|^2 \} = 1$. Here, $\mathbf s_0 \in \mathbb C^{N_t \times 1}$ represents a dedicated radar signal with covariance matrix $\mathbf V_0 \triangleq \mathbb{E}\{\mathbf s_0 \mathbf s_0^H\}$. The signals $\{s_l\}_{l=1}^L$ and $\mathbf s_0$ are assumed to be independent with each other.

For the receiver side, the complete model of the received signal at the FD BS can be expressed as
\begin{align}\label{signal:y}
\mathbf y^\text{BS} = \!\!\!\!\!\! \underbrace{\sum_{k=1}^K \mathbf h_k d_k}_\text{Communication signal}\!\!\!\!\!\! + \underbrace{ \beta_0 \mathbf A_0 \mathbf x}_\text{Target reflection} + \!\!\! \!\!\!\underbrace{\sum_{i=1}^I \beta_i \mathbf A_i \mathbf x}_\text{Echo signal of interferers} \!\!\!\!\!\! +  \underbrace{\mathbf H_\text{SI} \mathbf x}_\text{SI} + \mathbf n.
\end{align}
In the first term, $d_k \in \mathbb C$ denotes the signal transmitted from uplink user $k$, $k \in \{1,\cdots,K\}$, which satisfies $\mathbb E\{|d_k|^2\} = p_k$ with $p_k$ being the average transmit power of user $k$, and $\mathbf h_k \in \mathbb C^{N_r \times 1}$ represents the uplink channel.
The echo signal reflected by the target located at angle $\theta_0$ yields the second term, where $\beta_0 \in \mathbb C $ is the complex amplitude of the target and $\mathbf A_0 \triangleq \mathbf a_{r,0} \mathbf a_{t,0}^H $ denotes the two-way radar channel with $\mathbf a_{t,0}$ and $\mathbf a_{r,0}$ being the steering vectors towards angle $\theta_0$ of the transceiver antennas, respectively.
The third and the fourth terms are the signal-dependent interference, which correspond to the echoes from $I$ uncorrelated interferers located at angles $\{ \theta_i\}_{i=1}^I$ and $\theta_i \neq \theta_0,\ \forall i,$ and the residual self-interference (SI) after employing the SI cancellation techniques for ISAC systems \cite{M.Temiz2021TCCN} with $\mathbf H_\text{SI}$ being the residual SI channel, respectively. The last term, $\mathbf n$, stands for the additive white Gaussian noise (AWGN) with covariance $\sigma_r^2 \mathbf I_{N_r}$.

Denote the channel between downlink user $l$ and the BS by $\mathbf g_l \in \mathbb C^{N_t\times 1}$. The received signal at downlink user $l$ is then
\begin{align}\label{signal:yuser}
y^\text{User}_l = \mathbf g_l^H \mathbf v_l s_l + \sum_{l'=1,l'\neq l}^L \mathbf g_l^H \mathbf v_{l'} s_{l'} + \mathbf g_l^H \mathbf s_0 +\ n_l,\ \forall l,
\end{align}
where $n_l$ stands for the AWGN with variance $\sigma^2_l$.

\subsection{Radar and Communication SINR}
The performances of radar and communication systems highly depend on the corresponding SINRs.
We apply a receive beamformer $\mathbf u \in \mathbb C^{N_r \times 1}$ on $\mathbf y^\text{BS}$ to capture the desired target reflection. Based on (\ref{signal:y}), we obtain the radar SINR as
\begin{align}\label{sinr_r}
\gamma^\text{r}
= \frac{| \beta_0|^2 \mathbf u^H \mathbf A_0 \mathbf Q \mathbf A_0^H \mathbf u}
{\mathbf u^H ( \sum_{k=1}^K p_k  \mathbf h_k \mathbf h_k^H +  \mathbf B \mathbf Q {\mathbf B}^H + \sigma^2_r \mathbf I_{N_r} )\mathbf u},
\end{align}
where $\mathbf B \triangleq \sum_{i=1}^I \beta_i \mathbf A_i + \mathbf H_\text{SI}$ and
$
\mathbf Q \triangleq \mathbb E \{ \mathbf x \mathbf x^H\} =\sum_{l=1}^L \mathbf v_l \mathbf v_l^H + \mathbf V_0
$
denotes the covariance matrix of the downlink ISAC signal.
Similarly, by applying another set of receive beamformers $\{\mathbf w_k\}_{k=1}^K \in \mathbb C^{N_r \times 1}$ on $\mathbf y^\text{BS}$ to recover the data signals of the uplink users, we obtain the corresponding receive SINR of user $k$ by
\begin{align}\label{sinr_k}
\gamma_k^\text{c,UL}\!\!
=\!\! \frac{p_k \mathbf w_k^H \mathbf h_k \mathbf h_k^H \mathbf w_k}
{\mathbf w_k^H \!( \sum_{k'=1, k' \neq k}^K \! p_{k'}  \mathbf h_{k'} \mathbf h_{k'}^H \!\!\! +\!  \mathbf C \mathbf Q \mathbf C^H \!\! +\! \sigma_r^2 \mathbf I_{N_r} ) \mathbf w_k},\ \forall k,
\end{align}
where $\mathbf C \triangleq \sum_{i=0}^I \beta_i \mathbf A_i + \mathbf H_\text{SI}$.
As for the downlink communication, it follows from (\ref{signal:yuser}) that the SINR of downlink user $l$ is given by
\begin{align}\label{sinr_l}
\gamma_l^\text{c,DL} =& \frac{|\mathbf g_l^H \mathbf v_l |^2}{\sum_{l'=1,l'\neq l}^L |\mathbf g_l^H \mathbf v_{l'} |^2 + \mathbf g_l^H \mathbf V_0 \mathbf g_l  + \sigma^2_l},\ \forall l.
\end{align}

\subsection{Problem Formulation}
We aim at jointly optimizing the transmit power, $\{p_k\}_{k=1}^K$, at the uplink users, the receive beamformers, $\{ \mathbf w_k\}_{k=1}^K$ and $\mathbf u$, and the transmit beamforming, $\{\mathbf v_l\}_{l=1}^L$ and $\mathbf V_0$, at the BS,
so as to minimize the total transmit power consumption while guaranteeing the minimal SINR requirements of uplink communications, downlink communications, and radar sensing.
The corresponding problem is formulated as
\begin{align}\label{prob:minP}
\mathop \text{min} \limits_{\{ \mathbf w_k\}_{k=1}^K,\mathbf u, \mathbf V_0\succeq \mathbf 0,\atop \{\mathbf v_l\}_{l=1}^L, \{p_k \geq 0\}_{k=1}^K } \!
& \sum_{l=1}^L \| \mathbf v_l \|^2 + \text{Tr}(\mathbf V_0) + \sum_{k=1}^K p_k \nonumber \\
\text{s. t.}\quad\quad
& \gamma^\text{r} \!\geq\! \tau^\text{r}, \
 \gamma_k^\text{c,UL} \!\geq\! \tau_k^\text{c,UL}, \ \forall k, \
 \gamma_l^\text{c,DL} \!\geq\! \tau_l^\text{c,DL}, \ \forall l,
\end{align}
where $\tau^\text{r}$, $\tau_k^\text{c,UL}$, and $\tau_l^\text{c,DL}$ are the minimal SINR thresholds of sensing, uplink user $k$, and downlink user $l$, respectively.
Due to the nonconvexity and the coupled variables, it is generally hard to obtain the globally optimal solution of (\ref{prob:minP}).

\section{PROPOSED ALGORITHM}
\label{sec:3}
In this section, we handle (\ref{prob:minP}) by first determining the optimal receive beamformers in closed-form expressions. Then, we substitute them into (\ref{prob:minP}) and address the equivalent problem exploiting the SCA technique.

\subsection{Closed-Form Solutions to Receive Beamformers}
Note that the objective of (\ref{prob:minP}) does not depend on $\{\mathbf w_k\}_{k=1}^K$ and $\mathbf u$. To facilitate the fulfillment of the SINR constraints and reduce the power consumption, $\left\{\{\mathbf w_k\}_{k=1}^K, \mathbf u\right\}$ should be determined by maximizing the corresponding SINRs:
\begin{align}
\mathop \text{max}\limits_{ \mathbf u} \quad  \gamma^\text{r}, \quad
\mathop \text{max}\limits_{ \mathbf w_k} \quad  \gamma^\text{c,UL}_k , \ \forall k . \label{prob:w_k}
\end{align}

\begin{proposition}\label{prop:receiver}
The optimal solutions of (\ref{prob:w_k}) are respectively given by
$\mathbf u^* \!=\! ( \sum_{k=1}^K p_k \mathbf h_k \mathbf h_k^H \!+\! \mathbf B \mathbf Q \mathbf B^H \!+\! \sigma_r^2 \mathbf I_{N_r} )^{-1} \mathbf a_{r,0}$, and $
\mathbf w_k^* \!=\! (\sum_{k'=1, k'\neq k}^K p_{k'}  \mathbf h_{k'} \mathbf h_{k'}^H \!+\! \mathbf C \mathbf Q \mathbf C^H \!+\! \sigma_r^2 \mathbf I_{N_r} )^{-1} \mathbf h_k, \ \forall k.$
\end{proposition}
\textit{Proof.}
The problems in (\ref{prob:w_k}) belong to generalized Rayleigh quotient \cite{Rayleighquotient}. By invoking the results in \cite{Rayleighquotient}, we readily obtain the optimal receive beamformers. $\QEDB$

\subsection{Solutions to Transmit Beamforming and Power}
Substituting $\mathbf u^*$ and $\{\mathbf w_k^*\}_{k=1}^K$, $\gamma^\text{r}$ and $\gamma_k^\text{c,UL}$ become
$
\bar{\gamma}^\text{r}= |\beta_0|^2 \mathbf a_{t,0}^H \mathbf Q \mathbf a_{t,0} \mathbf a_{r,0}^H (\sum_{k=1}^K p_k  \mathbf h_k \mathbf h_k^H + \mathbf B \mathbf Q \mathbf B^H + \sigma_r^2 \mathbf I_{N_r} )^{-1} \mathbf a_{r,0}$ and
$
\bar{\gamma}_k^\text{c,UL}= p_k \mathbf h_k^H  (\sum_{k'=1, k'\neq k}^K p_{k'}  \mathbf h_{k'} \mathbf h_{k'}^H +  \mathbf C \mathbf Q \mathbf C^H + \sigma_r^2 \mathbf I_{N_r} )^{-1}\mathbf h_k, \ \forall k,$ respectively.
Now, (\ref{prob:minP}) relies on the transmit beamforming and power only.
To handle the complicated SINR constraints, i.e., $\bar{\gamma}^\text{r} \geq \tau^\text{r},
\bar{\gamma}_k^\text{c,UL} \geq \tau_k^\text{c,UL}, \ \forall k,$ and
$\gamma_l^\text{c,DL} \geq \tau_l^\text{c,DL}, \ \forall l,$
we introduce a set of auxiliary variables $\mathbf V_l \triangleq \mathbf v_l \mathbf v_l^H,\ \forall l$. With $\{\mathbf V_l\}_{l=1}^L$, we further define $\mathbf {\bar Q} \triangleq \sum_{l=1}^L \mathbf V_l + \mathbf V_0$, $\mathbf \Psi \triangleq \sum_{k=1}^K p_k  \mathbf h_k \mathbf h_k^H +  \mathbf B \mathbf {\bar Q} \mathbf B^H + \sigma_r^2 \mathbf I_{N_r}$, and $\mathbf \Phi_k
\triangleq \sum_{k'=1,k'\neq k}^K p_{k'}  \mathbf h_{k'} \mathbf h_{k'}^H +\mathbf C \mathbf {\bar Q} \mathbf C^H + \sigma_r^2 \mathbf I_{N_r}, \ \forall k,$ to ease notations.
After some straightforward algebraic operations, an equivalent reformulation of (\ref{prob:minP}) is written as
\begin{align}
\!\!\!\!\!\!\mathop \text{min} \limits_{\{\mathbf V_l \succeq \mathbf 0 \}_{l=0}^L, \atop \{p_k \geq 0 \}_{k=1}^K } \quad
\!\!\!\!& \sum_{l=0}^L \text{Tr}(\mathbf V_l)+ \sum_{k=1}^K p_k \label{prob:minPtildeRef}  \\
\!\!\!\!\!\!\text{s. t.}\quad\ \
& \mathbf a_{t,0}^H \mathbf {\bar Q} \mathbf a_{t,0} \mathbf a_{r,0}^H\mathbf \Psi^{-1} \mathbf a_{r,0} \geq \tau^\text{r}/|\beta_0|^2,                                 \tag{\ref{prob:minPtildeRef}{a}} \label{cons:minPrad} \\
& \mathbf h_k^H  \mathbf \Phi_k^{-1} \mathbf h_k \geq \tau_k^\text{c,UL}/ p_k, \ \forall k, \tag{\ref{prob:minPtildeRef}{b}} \label{cons:minPUL} \\
& (1\!+\!1/\tau_l^\text{c,DL})\mathbf g_l^H \mathbf V_l \mathbf g_l \!\geq\! \mathbf g_l^H \mathbf {\bar Q}\mathbf g_l\!+\! \sigma^2_l,  \ \forall l \geq 1.  \tag{\ref{prob:minPtildeRef}{c}} \label{cons:minPDL}
\end{align}
Note that we omitted the rank constraints of $\{\mathbf V_l\}_{l=1}^L$, i.e.,
$\text{rank}(\mathbf V_l) \leq 1, \ \forall l\geq 1,$
based on the rank relaxation technique \cite{GaussianR}.
The reformulation in (\ref{prob:minPtildeRef}) still has nonconvex constraints (\ref{cons:minPrad}) and (\ref{cons:minPUL}).

To obtain a more tractable form, we employ the iterative SCA technique to handle (\ref{cons:minPrad}) and (\ref{cons:minPUL}).
\begin{lemma}
With the SCA framework, a convex subset of (\ref{cons:minPrad}) is established as
\begin{align}\label{cvxApprox:rad}
f(\mathbf \Psi, \mathbf \Psi^{(i-1)}) \geq \tau^\text{r} ( \mathbf a_{t,0}^H \mathbf {\bar Q} \mathbf a_{t,0} )^{-1}/|\beta_0|^2,
\end{align}
where $f(\mathbf \Psi, \mathbf \Psi^{(i-1)}) \triangleq \mathbf a_{r,0}^H (\mathbf \Psi^{(i-1)})^{-1} \mathbf a_{r,0} - \mathbf a_{r,0}^H(\mathbf \Psi^{(i-1)})^{-1}\\
\times (\mathbf \Psi-\mathbf \Psi^{(i-1)})(\mathbf \Psi^{(i-1)})^{-1} \mathbf a_{r,0}$ with $\mathbf \Psi^{(i-1)}\triangleq \sum_{k=1}^K p_k^{(i-1)} \\ \times \mathbf h_k \mathbf h_k^H +  \mathbf B \mathbf {\bar Q}^{(i-1)} \mathbf B^H + \sigma_r^2 \mathbf I_{N_r}$ and $\mathbf {\bar Q}^{(i-1)} = \sum_{l=0}^L \mathbf V_l^{(i-1)}$. Here, $\{p_k^{(i-1)} \}_{k=1}^K$ and $\{\mathbf V_l^{(i-1)} \}_{l=0}^L$ denote the optimal solutions obtained in the $(i-1)$-th iteration.
Similarly, for (\ref{cons:minPUL}), a convex subset is given by
\begin{align}\label{cvxApprox:comk}
 f(\mathbf \Phi_k, \mathbf \Phi_k^{(i-1)}) \geq \tau_k^\text{c,UL}/ p_k,\ \forall k,
\end{align}
where $f(\mathbf \Phi_k, \mathbf \Phi_k^{(i-1)}) \triangleq  \mathbf h_k^H(\mathbf \Phi_k^{(i-1)})^{-1} \mathbf h_k - \mathbf h_k^H(\mathbf \Phi_k^{(i-1)})^{-1} \\
\times (\mathbf \Phi_k-\mathbf \Phi_k^{(i-1)}) (\mathbf \Phi_k^{(i-1)})^{-1} \mathbf h_k$, with $\mathbf \Phi_k^{(i-1)}\triangleq \sum_{k'=1,k'\neq k}^K p_{k'}^{(i-1)} \\ \times \mathbf h_{k'} \mathbf h_{k'}^H +  \mathbf C \mathbf {\bar Q}^{(i-1)} \mathbf C^H + \sigma_r^2 \mathbf I_{N_r}$ being calculated based on the solutions obtained in the $(i-1)$-th iteration.
\end{lemma}
\textit{Proof.} Proof can be found in the full paper version\cite{ourJSAC}. $\QEDB$

Exploiting the SCA technique, we address (\ref{prob:minPtildeRef}) by iteratively solving the following problem
\begin{align}\label{prob:minPSCA}
\mathop \text{min} \limits_{ \{\mathbf V_l \succeq \mathbf 0 \}_{l=0}^L, \atop \{p_k \geq 0 \}_{k=1}^K } \quad
& \sum_{l=0}^L \text{Tr}(\mathbf V_l) + \sum_{k=1}^K p_k \nonumber \\
\text{s. t.}\quad\quad
&\rm (\ref{cons:minPDL}),(\ref{cvxApprox:rad}), (\ref{cvxApprox:comk}).
\end{align}
This problem is convex and its globally optimal solution can be readily found via, e.g., the interior point method \cite{cvx}.
After solving (\ref{prob:minPSCA}), we update $\mathbf {\bar Q}^{(i)}$, $\mathbf \Psi^{(i)}$, and $\{ \mathbf \Phi_k^{(i)}\}_{k=1}^K$ exploiting the optimal solutions to $\{\{\mathbf V_l\}_{l=0}^L, \{p_k\}_{k=1}^K\}$ and then proceed to the $(i+1)$-th iteration.
Furthermore, according to \cite{SCAconvergence}, this iterative procedure converges to a Karush-Kuhn-Tucker (KKT) point of the problem in (\ref{prob:minPtildeRef}).

Upon convergence, we denote the obtained solution as $\{\{\mathbf {\widehat V}_l\}_{l=0}^L, \{\hat p_k \}_{k=1}^K\}$. Recall that $\mathbf V_l = \mathbf v_l\mathbf v_l^H,\ \forall l,$ and the rank-one constraints of $\{\mathbf {V}_l\}_{l=1}^L$ are omitted when solving (\ref{prob:minPtildeRef}).
To extract the beamforming vectors $\{\mathbf v_l\}_{l=1}^L$, i.e., the solution of (\ref{prob:minP}), we give the following theorem.

\begin{theorem}\label{theorem:rank1soltuion}
A solution of (\ref{prob:minPtildeRef}) achieving the same power consumption as $(\{\mathbf {\widehat V}_l\}_{l=0}^L, \{\hat p_k \}_{k=1}^K)$ while satisfying the relaxed rank-one constraints can be constructed as
\begin{align}
\mathbf {V}^*_l =&\ \mathbf {v}^*_l ({\mathbf v}_l^*)^H, \ \forall l\geq 1,\quad p_k^* = \hat p_k, \ \forall k \nonumber\\
\mathbf {V}_0^* =&\ \sum_{l=1}^L \mathbf {\widehat V}_l + \mathbf {\widehat V}_0 - \sum_{l=1}^L \mathbf {v}_l^* (\mathbf {v}^*_l)^H,  \label{def:tildeV0}
\end{align}
where $\mathbf {v}^*_l = (\mathbf g_l^H \mathbf {\widehat V}_l \mathbf g_l )^{-1/2} \mathbf {\widehat V}_l \mathbf g_l, \ \forall l \geq 1.$
\end{theorem}
\textit{Proof.}
Proof can be found in \cite{ourJSAC}. $\QEDB$

Theorem \ref{theorem:rank1soltuion} admits a new solution of (\ref{prob:minPtildeRef}), which
satisfies the relaxed rank-one constraints and attains the same performance as a KKT solution (i.e., $\{\{\mathbf {\widehat V}_l\}_{l=0}^L, \{\hat p_k \}_{k=1}^K\}$). Based on $\{\{\mathbf {V}^*_l\}_{l=0}^L, \{p_k^* \}_{k=1}^K\}$, we thus can recover the solution of (\ref{prob:minP}) as $\{\{\mathbf {v}_l^*\}_{l=1}^L, \mathbf {V}_0^*,\{p_k^* \}_{k=1}^K, \mathbf u^*, \{\mathbf w_k^*\}_{k=1}^K\}$.

The procedure for solving (\ref{prob:minP}) is summarized as Algorithm~\ref{alg:minP}.
The main computational burden stems from solving (\ref{prob:minPSCA}) in each iteration, whose computational complexity, according to \cite{complexity}, is given by $\mathcal O\left( N_t^{6.5} L^{3.5} + N_t^{4} L^{2} K^{1.5} \right)$ .

\begin{algorithm}[t]
\caption{Proposed Algorithm for solving (\ref{prob:minP})}
\label{alg:minP}
\begin{algorithmic}[1]
\STATE Initialize $\{\{\!\mathbf V_l^{(0)}\!\}_{l=0}^L, \{\!p_k^{(0)}\! \}_{k=1}^K\}$ and iteration index $i\! =\! 0$.\REPEAT
    \STATE Set $i = i + 1$.\\
    \STATE
    Solve (\ref{prob:minPSCA}) with $\{\{\mathbf V_l^{(i-1)}\}_{l=0}^L, \{p_k^{(i-1)} \}_{k=1}^K\}$ and update $\{\{\mathbf V_l^{(i)}\}_{l=0}^L, \{p_k^{(i)} \}_{k=1}^K\}$.\\

\UNTIL \textit{Convergence}.
\STATE Calculate $\{\{\mathbf {v}_l^*\}_{l=1}^L, \mathbf {V}_0^*,\{p_k^* \}_{k=1}^K\}$ according to (\ref{def:tildeV0}).
\STATE Calculate $\{\mathbf u^*, \{\mathbf w_k\}_{k=1}^K\}$ according to Proposition 1.
\end{algorithmic}
\end{algorithm}

\section{NUMERICAL RESULTS}
\label{sec:4}
We present numerical simulations to evaluate the performance
of the proposed algorithm.

We set $N_t = N_r = 8$ and $K=L=3$. All the user channels are assumed to follow the Rayleigh fading model with a path loss of $-99$~dB.
The target and $I=2$ interferers are located at $\theta_0 = 0^{\circ}$, $\theta_1 = -60^{\circ}$, and $\theta_2 = 45^{\circ}$, respectively, with the channel power gains of $| \beta_0|^2 = -100$~dBm and $| \beta_1|^2 = | \beta_2|^2 = -90$~dBm.
We set the noise power as $\sigma^2_r=\sigma^2_l= -100$~dBm, $\forall l$. For the residual SI channel, we let $\mathbf H_\text{SI}\triangleq \alpha_\text{SI}\mathbf{\tilde H}_\text{SI}$ with setting $\alpha_\text{SI} = -110$~dB and modeling each entry of $\mathbf{\tilde H}_\text{SI}$ as a unit-modulus variable with random phase \cite{M.Temiz2021TCCN}.
The required SINR thresholds are set to $\tau^\text{c,DL}_l = 8$ dB, $\forall l$, $ \tau^\text{c,UL}_k = 5$ dB, $\forall k$, and $\tau^\text{r} = 6$ dB, respectively.

\begin{figure}[t]
\centering
\subfigure[Beampattern gain]{
\begin{minipage}[t]{0.5\linewidth}
\centering
\includegraphics[width=1.7in]{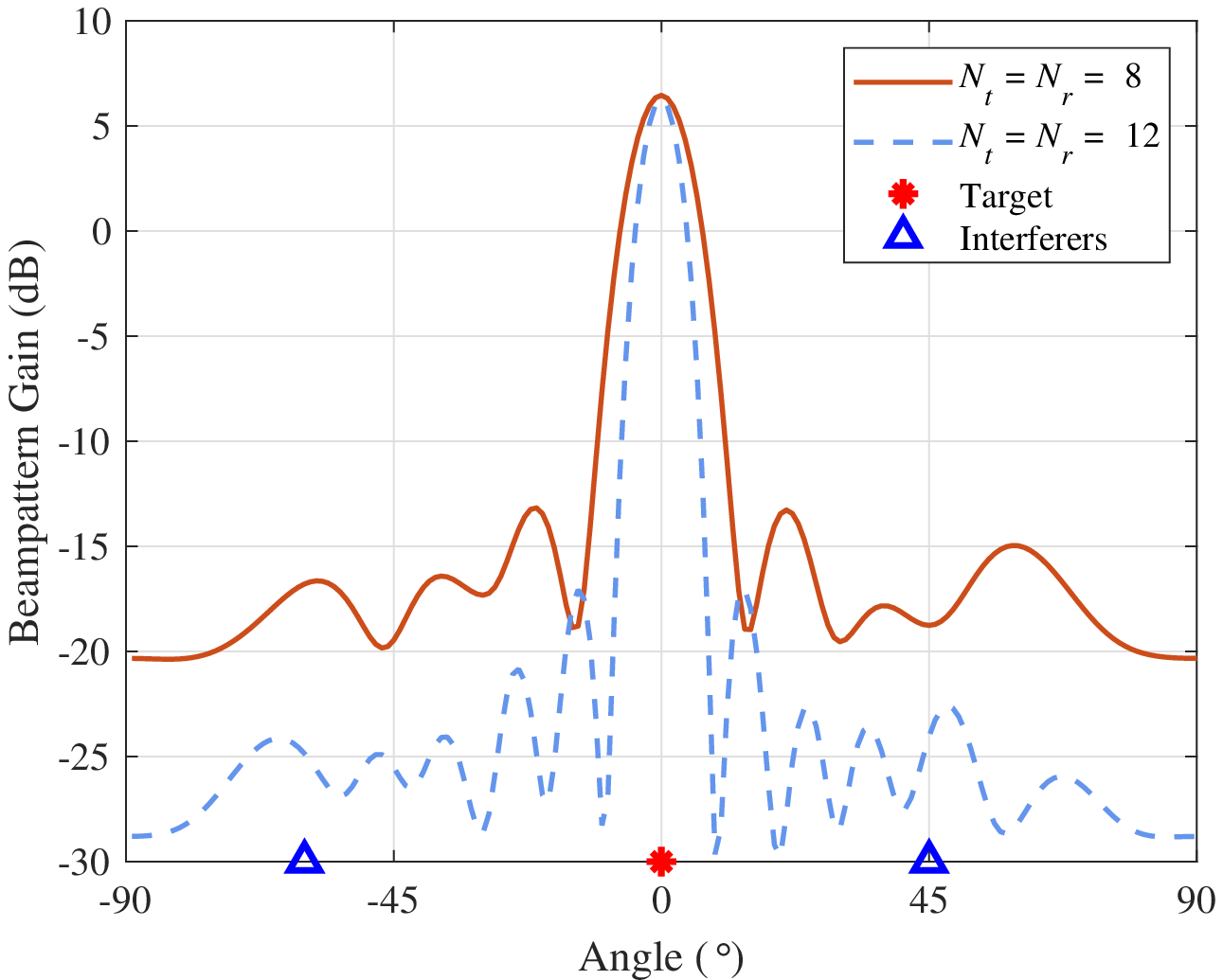}
\end{minipage}%
}%
\subfigure[Power consumption]{
\begin{minipage}[t]{0.5\linewidth}
\centering
\includegraphics[width=1.7in]{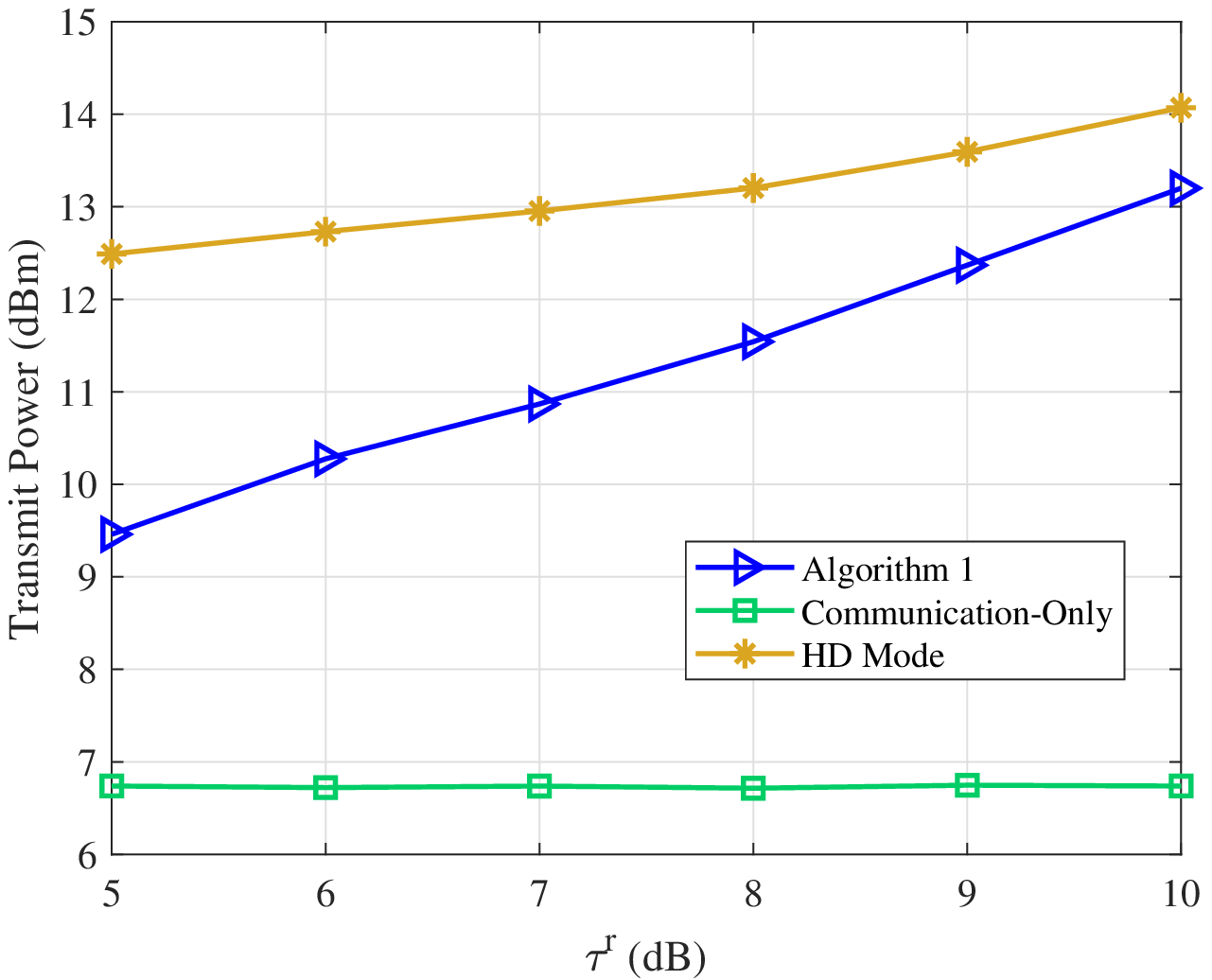}
\end{minipage}%
}%
\centering
\caption{Performance achieved by Algorithm \ref{alg:minP}.}\label{fig:1}
\end{figure}

%

In Fig. \ref{fig:1}(a), we show the beampattern gain achieved by Algorithm~\ref{alg:minP}, which is defined by $\frac{|(\mathbf u^*)^H \mathbf A_\theta \mathbf x^*|^2}{\sigma^2_r (\mathbf u^*)^H \mathbf u^*}$, where $\mathbf A_\theta = \mathbf a_{r,\theta} \mathbf a_{t,\theta}^H$ with $\theta$ varying in $[-90^{\circ},90^{\circ}]$ and $\mathbf x^*$ represents the optimized transmit signal. Obviously, it is seen that the main beam is allocated to the target at direction $\theta_0 = 0^{\circ}$ and two relatively deep nulls are placed towards the interferers. This indicates the effectiveness of the proposed algorithm.
Moreover, when the number of the transceiver antennas increases from 8 to 12, lower power consumption and more precise beampattern can be constructed due to additional degrees-of-freedoms brought by the larger antenna arrays.

Fig. \ref{fig:1}(b) shows the minimum total power versus the radar SINR threshold $\tau^\text{r}$.
In the figure, the communication-only scheme omits the sensing constraint when solving (\ref{prob:minP}). The HD model scheme separates downlink and uplink communications occupying two slots while the downlink sensing is continuously performed at the BS for achieving high-accuracy radar sensing (see \cite{ourJSAC} for details).
Observe that the power consumed by the communication-only design remains unchanged since it does not contain the sensing constraint. On the contrary, when $\tau^\text{r}$ increases, the power consumption of Algorithm \ref{alg:minP} and the HD mode and their performance gaps between the communication-only design are enlarged due to the hasher requirement for radar sensing.
Moreover, compared to the conventional HD mode, Algorithm~\ref{alg:minP} yields a much lower power consumption, which validates the superiority of the proposed FD scheme.

\section{CONCLUSION}
\label{sec:5}
In this paper, we studied the joint optimization of an FD communication-based ISAC system for transmit power minimization.
We first derived the optimal uplink receive beamformers in closed forms with respect to the downlink transmit beamforming and the uplink power. Then, we developed an iterative algorithm to solve the remaining problem.
Simulation results verified the effectiveness of the proposed algorithm and showed the tremendous advantage of our considered FD communication-based ISAC system over the previous frameworks that integrated sensing with HD communication.

\vfill\pagebreak

\bibliographystyle{IEEEbib}
\bibliography{strings}

\end{document}